\documentclass[12pt]{iopart}
\usepackage{graphicx}

\begin{document}

\title{Controlling crystal symmetries in phase-field crystal models}

\author{Kuo-An Wu$^1$, Mathis Plapp$^2$ and Peter W Voorhees$^{1, 3}$}

\address{$^1$ Department of Materials Science and Engineering, Northwestern University, Evanston, Illinois 60208, USA}
\address{$^2$ Physique de la Mati\`ere Condens\'ee, 
\'Ecole Polytechnique, CNRS, 91128 Palaiseau, France}
\address{$^3$ Department of Engineering Sciences and Applied Mathematics, Northwestern University, Evanston, Illinois 60208, USA}
\ead{\mailto{mathis.plapp@polytechnique.fr}, \mailto{kuoan-wu@northwestern.edu}, \mailto{p-voorhees@northwestern.edu}}

\begin{abstract}

We investigate the possibility to control the symmetry of
ordered states in phase-field crystal models by tuning
nonlinear resonances. In two dimensions, we find that a
state of square symmetry as well as coexistence between
squares and hexagons can be easily obtained. In contrast,
it is delicate to obtain coexistence of squares and liquid.
We develop a general method for constructing free energy 
functionals that exhibit solid-liquid coexistence with 
desired crystal symmetries. As an example, we develop 
a free energy functional for square-liquid coexistence 
in two dimensions. A systematic analysis for determining 
the parameters of the necessary nonlinear terms is provided.
The implications of our findings for simulations of 
materials with simple cubic symmetry are discussed.
\end{abstract}

\pacs{68.08.-p,45.70.Qj, 47.54.-r, 81.16.Rf }


\section{Introduction}

The phase-field crystal (PFC) model has rapidly gained popularity
in recent years as a potentially powerful tool to simulate the
evolution of materials with atomistic resolution on time
scales that are orders of magnitude larger than those that can
be attained by molecular dynamics simulations. While it was
originally \cite{Elder02,Elder04} inspired by phenomenological 
equations developed in the general context of pattern 
formation \cite{SH,Cross93}, it has now been established
\cite{Elder07,Wu07,SvT09} that it can be obtained as a 
simplification of classical density functional theory 
(DFT) of freezing \cite{Ramakrishnan79,Singh91}. 

To make the PFC model useful for practical applications in
materials science, it has to be established how the model
parameters need to be chosen in order to reproduce the physical
properties of a given material as closely as possible.
This is currently a very active area of research, and several
contributions of the present volume are dedicated to this
subject. One of the most fundamental properties of a material
is of course its crystal structure. As detailed below, the original 
PFC model \cite{Elder02,Elder04} contains only two parameters:
a scaled global density $\bar\psi$ and a scaled temperature 
$\epsilon$. The latter indicates the distance to the critical 
point of the PFC model, which is located at $\bar\psi=\epsilon=0$.
For small $\epsilon$, the PFC model exhibits, besides the 
unstructured liquid phase, periodic solutions of nematic (stripe) 
and hexagonal symmetry in two dimensions, and of bcc symmetry in 
three dimensions. Recently, it was found that for larger values
of $\epsilon$, fcc and hcp structure also exist in three
dimensions \cite{Tegze09,Toth10,Jaatinen10}. However, the
parameter $\epsilon$ also controls several other properties
of the model, such as the width of the solid-liquid interfaces 
and their interfacial free energy. Since, to match a given
material, these quantities need to be adjusted independently,
more degrees of freedom are needed in the model. 

It turns out that it is not straightforward to obtain 
periodic ground states with symmetries that differ from 
the ``natural'' ones. This is due to the physics which 
governs the formation of periodic states in the PFC
model: the liquid state is unstable with respect to the
formation of periodic solutions if their wavelength falls
within a narrow band centred around a characteristic 
length scale. The periodic density patterns that correspond
to crystalline phases are stabilised by nonlinear terms,
which lead to the interaction (resonance) of density waves with 
different unstable wave vectors. The hexagon/bcc and stripe 
patterns arise from the simplest nonlinearities involving triadic
and quartic resonances of wave vectors with equal modulus, 
respectively, which explains their ubiquity in nature \cite{Cross93}.
This also naturally explains why new structures appear for larger
values of $\epsilon$ \cite{Tegze09,Toth10,Jaatinen10}: since
the range of unstable wavelengths increases with $\epsilon$, new
resonances become possible.

A way to control the selection of crystal structures is thus to modify
the nonlinear resonances. Two ways to do this have been explored
in the literature, both inspired by earlier work in the physics
of pattern formation. The first idea is to make {\em two} bands
of wavelengths unstable, and to use the ratio between the two
characteristic wavelengths as an additional geometric parameter which
allows to stabilise patterns of a certain symmetry \cite{Lifshitz97}.
Using this approach, PFC models with a ground state of fcc
symmetry were recently developed \cite{Wu10,Greenwood10}. While these 
models are robust and versatile, they have two potential drawbacks, 
namely (i) in the analogy with DFT, the two distinct length scales
should arise from two distinct peaks in the liquid structure
factor; for the simplest version of the model, a structure
factor that is very different from those typically measured
has to be used, and (ii) the equation of motion for the density 
field contains spatial derivatives of up to 10th order, which 
makes simulations in geometries that cannot easily be handled 
in reciprocal space very cumbersome.

The second approach, which will be our main focus, is to add
new nonlinear terms to the model which modify the strength
of the resonances and can thus favour patterns of different
symmetries. Here, we will restrict our investigations to
two dimensions, where the obvious missing state is a square
pattern. The question of how a square pattern can arise
from a rotationally invariant homogeneous state has been
extensively studied in the context of convection patterns \cite{Busse78},
because squares can be observed under certain experimental
conditions. It has been demonstrated that simple additional
nonlinearities can stabilise squares \cite{Gertsberg81,Bestehorn84},
and that squares can coexist with hexagons \cite{Herrero94,Kubstrup96}.
These phenomenological equations have already been used to study 
the transition from hexagonal to square symmetry upon a change 
of the control parameters, in a spirit very close to the one of
the PFC model, before that name was actually 
coined \cite{Matsushita98,Enomoto01}. The geometric properties
of grain boundaries between two domains of square symmetries
have also been studied \cite{Boyer05}. Finally, isolated
patches of squares in coexistence with the unstructured
state (``oscillons'') have also been found in similar 
equations \cite{Sakaguchi97,Crawford99}.

Here, we investigate whether it is possible to use this approach
to design a simple and robust PFC model with square symmetry. 
We find that it is indeed straightforward to obtain squares
as well as coexistence of squares and hexagons following the
recipes found in the literature. In contrast, square-liquid
coexistence can only be obtained with a combination of several
nonlinear terms and carefully tuned parameters. We develop a
systematic approach to determine suitable parameter ranges
that yield stable square-liquid interfaces, and confirm our
analytic calculations by numerical simulations. We also evaluate
the elastic coefficients of the square solid for the parameters
used in the simulations.

The remainder of the paper is organised as follows. In Section
\ref{sec:model}, we define the model and recall the calculation
of nonlinear resonances and the resulting selection of symmetry.
In Section \ref{sec:states} we calculate the free energies of
the various periodic patterns in a generalised PFC model in
two dimensions, with particular emphasis on square-liquid coexistence. 
Section \ref{sec:numerics} presents the comparison between our 
analytic results and numerical calculations; the elastic coefficients
are evaluated in Section \ref{sec:elastic}, and Section
\ref{sec:conclusions} gives a brief conclusion.

\section{The model}
\label{sec:model}

\subsection{General considerations}
For the purpose of exposition, we start with the standard
PFC model \cite{Elder02,Elder04}, which has the dimensionless
free energy functional 
\begin{eqnarray}
F = \int \rmd\vec{r} \left\{
{\psi\over 2} [-\epsilon + (\nabla^2 + 1)^2] \psi + {\psi^4 \over 4}
\right\},
\label{energy}
\end{eqnarray}
where $\psi$ is a dimensionless particle density measured 
from a constant reference value, and $\epsilon$
is a constant. This form of the free energy functional was 
originally proposed by Swift and Hohenberg as a phenomenological 
description of patterns that 
emerge in various hydrodynamical systems \cite{SH}. In this
context, $\epsilon$ is the distance from the bifurcation
threshold at which the unstructured (quiescent) state becomes
unstable. We have used dimensionless units in which
the modulus of the characteristic wave number is unity, and
energy and time have been scaled by appropriate quantities as
detailed in \cite{Wu07}. The manner by which this free energy functional
and its parameters can be related to classical DFT is detailed
in \cite{Elder07,Wu07,SvT09} and need not be repeated here.
 
In the original Swift-Hohenberg equation, $\psi$ is a linear
combination of the deviations of the temperature and velocity
fields from the quiescent state. It is treated as a non-conserved 
order parameter, and thus the system is free to evolve towards 
the global free energy minimum. In contrast, in the PFC model the 
field $\psi$ is a local particle density and hence a conserved quantity.
This provides an additional control parameter, namely, the average 
density $\bar\psi \equiv \int \psi d\vec{r}  / \int d\vec{r}$.

The free energy of periodic states can be evaluated using the
one-mode approximation, in which the density field is given 
by a sum of density waves,
\begin{eqnarray}
\psi(\vec{r}) = \bar\psi+\delta\psi(\vec{r})
              = \bar\psi + \sum_j A_{\vec{K}_j} \exp(\rmi \vec{K}_j \cdot \vec{r})
              = \bar\psi + A \sum_j  \exp(\rmi \vec{K}_j \cdot \vec{r}),
\label{one-mode}
\end{eqnarray}
where $\vec{K}_j$ are the principal reciprocal lattice vectors
of the considered structure (with $|\vec{K}_j|\equiv q$), and $A$
is the amplitude of the density waves, where we have used the 
fact that for a homogeneous solid all density waves have the 
same amplitude, $|A_{\vec{K}_j}| = A$.
This ansatz is inserted into the free energy functional, and
the result is integrated over one unit cell. In this procedure,
all terms that contain oscillatory exponential factors integrate out, 
and only the products of exponentials in which two, three, or four 
of the $\vec{K}_j$'s sum up to zero contribute to the final
result. Thus, non-trivial contributions arise from the triadic and
quartic resonances. Whereas the latter can generate stripes 
(or squares, see below), hexagonal and bcc ordered phases are 
stabilised by triadic resonances because the principal reciprocal 
lattice vectors are able to form ``triads'' (i.e., closed triangles,
for example, $\langle 110\rangle$, $\langle \bar{1} 0 1\rangle$ 
and $\langle 0 \bar{1} \bar{1}\rangle$ for bcc lattices). The 
cubic terms that are needed for triadic resonances to occur 
are generated by the expansion of the $\psi^4$ term: 
$(\bar\psi+\delta\psi)^4=(\delta\psi)^4+4\bar\psi(\delta\psi)^3+\ldots$ .
Therefore, they are absent for $\bar\psi=0$ and become increasingly
important with increasing $|\bar\psi|$. This explains the sequence 
of phases found in the phase diagram of the PFC model \cite{Elder04}
with increasing $|\bar\psi|$: rolls (stripes) $\to$ hexagons $\to$
liquid in two dimensions, and rolls $\to$ hexagons $\to$ bcc $\to$
liquid in three dimensions. In addition, since the free energy
expressed in terms of $A$ contains terms in $A^2$, $A^3$, and
$A^4$, the bifurcation from liquid to hexagons can be transcritical,
and a first-order transition from liquid to hexagons/bcc with a
finite coexistence zone in $\bar\psi$ is possible. In contrast, 
for crystals with square, simple-cubic and face-centred-cubic 
lattices, no triads can be formed, and the free energy contains only
terms in $A^2$ and $A^4$. This makes the transition from liquid
to squares supercritical, and it was shown by weakly nonlinear 
analysis \cite{Wu07, Wu10} that solid-liquid coexistence is
impossible under these conditions.

From the above considerations, we can conclude that the
model has to fulfil three requirements in order to obtain
square-liquid coexistence. First, the quartic 
resonances need to favour squares rather than stripes, second
the transition from squares to liquid must be subcritical to
give rise to solid-liquid coexistence, and third the square
state must have a lower free energy than hexagons even in
the presence of triadic resonances that occur for 
$\bar\psi\ne 0$. We will address these questions in the
following, and we will develop a systematic approach 
to determine coefficients that favour periodic states with 
cubic symmetry.

\subsection{Anisotropic terms}
Let us first discuss nonlinear terms that can be used to favour squares 
in two dimensions. To this end, it is useful to consider rhombi,
that is, a density field composed just of two sets of density 
waves with wave vectors $\pm\vec{K}_1$ and $\pm\vec{K}_2$,
which have equal magnitude $q$ and form an arbitrary angle 
$\theta$ between them,
\begin{eqnarray}
\delta \psi(\vec{r}) = 
A\; [ \exp(\rmi \vec{K}_1 \cdot \vec{r}) + \exp(\rmi \vec{K}_2 \cdot \vec{r}) + c.c.].
\label{rhombi}
\end{eqnarray}
Obviously, there are no triadic resonances. The quartic resonances
generated by the $\psi^4$ term in equation (\ref{energy}) give
a result that is independent of $\theta$. Thus, this term does
not favour any particular symmetry.

\begin{figure}
\begin{center}
\begin{minipage}[b]{10cm}
\includegraphics[width=0.95\textwidth, angle=0]{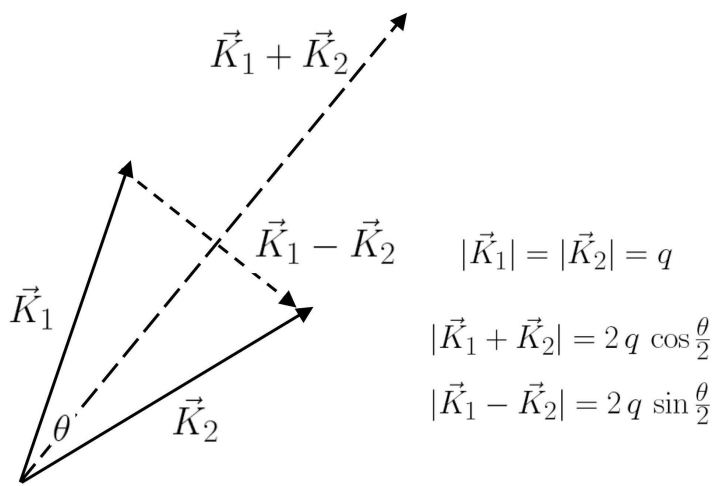}
\end{minipage}
\vskip 0.2cm
\caption{Terms of the form $\psi^2 \Delta^n \psi^2$ give rise 
to a $\theta$-dependent free energy through the above vector 
addition operations.\label{fig:vectors}}
\end{center}
\end{figure}

One possibility of an ``anisotropic'' term, used in the literature 
\cite{Bestehorn84,Herrero94,Kubstrup96}, is a term proportional
to $\psi^2 \Delta^2 \psi^2$, where $\Delta=\nabla^2$ is the
Laplace operator. Here, we consider a more general set of nonlinear 
terms of the form $g_{2n} \, \psi^2 \Delta^n \psi^2$ in the 
free energy functional, where $g_{2n}$ are constant coefficients.
Following the procedure outlined above, we obtain 
\begin{eqnarray}
\fl
g_{2n} \int \rmd \vec{r} \,\, \psi^2 \Delta^n \psi^2 =  
g_{2n}  \int \rmd \vec{r} \,\,
& & 
\left( 
{\bar\psi}^2 \Delta^n ( \delta\psi )^2 + 4 {\bar\psi}^2 \delta\psi \Delta^n \delta \psi 
+ 2\bar\psi \delta \psi \Delta^n ( \delta \psi )^2 
\right. \nonumber \\
& & 
\left. + 2\bar\psi ( \delta \psi )^2 \Delta^n \delta\psi + ( \delta\psi )^2 \Delta^n ( \delta \psi )^2
\right).
\label{nonlinear}
\end{eqnarray}
Substituting equation~(\ref{rhombi}) into equation~(\ref{nonlinear}) yields
\begin{eqnarray}
\fl
{g_{2n} \over V }\int \rmd \vec{r} \,\, \psi^2 \Delta^n \psi^2 =
& &
(-1)^n g_{2n} \left[
8 {\bar\psi}^2 A^2 \left(
|\vec{K}_1|^{2n} + |\vec{K}_2|^{2n}
\right)
\right. \nonumber \\
& &
\left.
+ 2A^4  \left(
  |2 \vec{K}_1|^{2n} +  |2 \vec{K}_2|^{2n}
+ 4 |\vec{K}_1+\vec{K}_2|^{2n}
+ 4 |\vec{K}_1-\vec{K}_2|^{2n}
\right)
\right],
\label{theta_dep}
\end{eqnarray}
where $V=\int d\vec{r}$ is the volume of one unit cell. Together with the 
vector addition properties illustrated in Figure~\ref{fig:vectors}, 
\begin{eqnarray} 
|\vec{K}_1 + \vec{K}_2| &=& 2q \cos{\theta \over 2} \nonumber \\ 
|\vec{K}_1 - \vec{K}_2| &=& 2q \sin{\theta \over 2}, 
\end{eqnarray} 
equation~(\ref{theta_dep}) can be rewritten as
\begin{eqnarray}
\fl
{g_{2n} \over V }\int \rmd \vec{r} \,\, \psi^2 \Delta^n \psi^2 &=& 
(-1)^n g_{2n} q^{2n} \left[ \left( 16 {\bar\psi}^2 A^2 + 
2^{2n+2} A^4
\right) + 8 A^4 \left(
\cos^{2n}{\theta \over 2}
+\sin^{2n}{\theta \over 2}
\right)
\right] 
\nonumber \\
&=& 
 f_{\rm isotropic} + f_{\rm anisotropic}.
\end{eqnarray}
where
\begin{eqnarray}
f_{\rm isotropic} \equiv (-1)^n g_{2n} q^{2n} \left( 16 {\bar\psi}^2 A^2 +  
2^{2n+2} A^4 
\right)
\end{eqnarray}
is independent of the angle $\theta$, but depends on $\bar\psi$,
and the anisotropic part of the free energy that depends on the
angle and thus on the symmetry of the pattern is
\begin{eqnarray}
f_{\rm anisotropic} \equiv 8 (-1)^n g_{2n} q^{2n} A^4
\left( 
\cos^{2n}{\theta \over 2} 
+\sin^{2n}{\theta \over 2} 
\right).
\label{anisotropy}
\end{eqnarray}
For $n=1$, the dependence on $\theta$ in the anisotropic part 
vanishes because of the trigonometric identity
$\cos^2(\theta/2) + \sin^2(\theta/2)=1$. However, for $n=2$, 
the anisotropic part of the free energy is a simple function 
of $\theta$ which is minimised by $\theta = \pi/2$ if $g_4 >0$ 
and by $\theta = 0$ if $g_4 <0$. Similar results are obtained 
for $n=3$; however, the square symmetry is favoured when $g_6 <0$.
It is interesting to note that for $n \geq 4$, the anisotropic 
part of the free energy contains higher order harmonics, which 
suggests a possible means to favour arbitrary angles between
density waves by an appropriate combination of such terms.

Another type of nonlinearity used in the literature 
\cite{Gertsberg81,Matsushita98,Enomoto01} to favour squares
is $|\nabla\psi|^4$. Indeed, repeating the above calculation 
for rhombi with a term of the form $s_4|\nabla\psi|^4/4$
with $s_4$ a constant, we find
\begin{eqnarray}
{s_4 \over V }\int \rmd \vec{r} \,\,\frac 14 |\nabla\psi|^4 &=&
  s_4q^4A^4(5+\cos^2\theta)
\end{eqnarray}
which can again be split in an isotropic and an anisotropic part.
The anisotropic part is minimised for $\theta=\pi/2$ if $s_4>0$.
Note that, in this case, the isotropic part does not depend 
on $\bar\psi$.

The above calculations are valid both in two and three dimensions.
In three dimensions, an angle of $\pi/2$ between density waves
corresponds to a simple cubic structure, which can therefore
be obtained in a straightforward manner using this method.

\subsection{Subcritical bifurcation}
For crystal structures without triadic interactions, since to lowest
order the free energy can be expressed in even powers of $A$, a subcritical 
bifurcation is required for solid-liquid coexistence. The generic 
form of a free energy that exhibits a subcritical bifurcation is 
\begin{eqnarray}
f = \alpha A^2 - \beta A^4 + \gamma A^6
\label{f246}
\end{eqnarray}
where $\beta$ and $\gamma$ are positive. 
The parameter $\alpha$ controls the growth rate of small 
perturbations of the liquid state (linear stability); the liquid is
stable for $\alpha>0$. The $\psi^4$ term in the standard 
PFC free energy functional shown in equation~(\ref{energy}) 
results in a $\bar\psi$-dependent quadratic coefficient.
Within the one-mode approximation, for this model
$\alpha\sim -\epsilon + (1-q^2)^2 + 3\bar\psi^2$ for 
density waves with wave number $q$.

All nonlinear terms discussed so far yield contributions up to 
order $A^4$. To obtain the necessary contribution in $A^6$, some
higher order nonlinearity needs to be added. An obvious possibility
would be just $\psi^6$, which was indeed used in \cite{Sakaguchi97,Crawford99}
to model oscillons. The problem with this choice is that 
for $\bar\psi\neq 0$ it generates additional cubic and quintic 
terms in the free energy that tend to lower the free energy for 
hexagonal and bcc phases. As a consequence, this nonlinearity
tends to generate the sequence squares $\to$ hexagons $\to$ liquid 
in the phase diagram; this was the case for all the cases that we 
have investigated.

An alternative choice, which will be made in the following, is
to add $s_6 |\nabla \psi|^6/6$ to the free energy, which generates 
terms of order $A^6$, but no additional terms depending on $\bar\psi$.
As will be shown below, it is indeed possible to generate a subcritical
bifurcation that favours squares by using both $s_4  |\nabla \psi|^4/4$ 
and $s_6 |\nabla \psi|^6/6$ with $s_4 <0$ and $s_6 >0$. However, the
resulting square-liquid equilibrium is metastable, whereas the
state of lowest energy are hexagons. In order to be able to adjust 
the respective energies of these states, we must also include 
the nonlinear terms of the form $\psi^2 \Delta^2 \psi^2$ and 
$\psi^2 \Delta^3 \psi^2$ discussed previously. The procedure
how to choose appropriate coefficients for all these terms
will be detailed in the next section.

\section{Selection of crystal symmetry}
\label{sec:states}

\subsection{Free energies}
The free energy functional considered in the following is 
\begin{eqnarray}
F = \int \rmd \vec{r} \left( \right. & & \left. 
{\psi \over 2} \left[
-\epsilon + (\nabla^2 + 1)^2 
\right] \psi + {\psi^4 \over 4}
+ {g_{4} \over 4} \psi^2 \Delta^2 \psi^2 +
{g_{6} \over 4} \psi^2 \Delta^3 \psi^2 +
\right. \nonumber \\
& &
\left.
{s_4 \over 4} |\nabla \psi|^4
+{s_6 \over 6} |\nabla \psi|^6 \right).
\label{F_sub_cubic}
\end{eqnarray}
In two dimensions, the competing patterns are rolls, squares and 
hexagons. The corresponding density fields in the one-mode 
approximation of equation~(\ref{one-mode}) are
\begin{eqnarray}
\psi_{\rm roll}(\vec{r}) &=& \bar\psi + 2A \cos(qx), \nonumber \\
\psi_{\rm square}(\vec{r}) &=& \bar\psi + 2A \left( \cos{qx} + \cos{qy} \right), \nonumber \\
\psi_{\rm hex}(\vec{r}) &=& \bar\psi + 2A \left( 2\cos{qx\over 2}\cos{\sqrt{3}qy \over 2} + \cos{qx} \right),
\label{ansatz}
\end{eqnarray}
where we recall that $q$ is the magnitude of the principal reciprocal 
lattice vectors. The free energy is computed by substituting 
equation~(\ref{ansatz}) into equation~(\ref{F_sub_cubic}) and 
integrating over one unit cell, which yields
\begin{eqnarray}
f_{\rm roll} = f_L &+& \left(
-\epsilon + (1-q^2)^2 + (3 + 2g_4 q^4 - 2 g_6 q^6) \bar\psi^2
\right) A^2 \nonumber \\
&-& {1\over 2} \left(
-3 -3 s_4 q^4 -16 g_4 q^4 + 64 g_6 q^6
\right) A^4 + {10 \over 3} s_6 q^6 A^6,
\label{roll_energy}
\end{eqnarray}
\begin{eqnarray}
f_{\rm square} = f_L &+& 2 \left(
-\epsilon + (1-q^2)^2 + (3 + 2g_4 q^4 - 2 g_6 q^6) \bar\psi^2
\right) A^2 \nonumber \\
&-&  \left(
-9 -5 s_4 q^4 -32 g_4 q^4 + 96 g_6 q^6
\right) A^4 + {56 \over 3} s_6 q^6 A^6,
\label{square_energy}
\end{eqnarray}
\begin{eqnarray}
f_{\rm hex} = f_L &+& 3 \left(
-\epsilon + (1-q^2)^2 + (3 + 2g_4 q^4 - 2 g_6 q^6) \bar\psi^2
\right) A^2 \nonumber \\ 
&+& 12 \bar\psi \left(
1 + g_4 q^4 - g_6 q^6
\right) A^3
\nonumber \\
&-&  \left(
-{45 \over 2} -{27 \over 2} s_4 q^4 -84 g_4 q^4 + 264 g_6 q^6
\right) A^4 + {91} s_6 q^6 A^6,
\label{hex_energy}
\end{eqnarray}
where $f$ is the free energy density, $f\equiv F / V$, and 
$f_L = (-\epsilon+1) \bar\psi^2/2 + \bar\psi^4/4$ is the free
energy density of the liquid. The values of $A$ and $q$ are 
determined by minimising the above free energy with respect 
to $A$ and $q$. One important difference with respect to the
standard PFC model should be noted. In the latter, $q$ 
appears only in the gradient term. Therefore, $q$ and $A$
decouple, and the value of $q$ that minimises the free 
energy is always equal to unity. In contrast, the new terms 
introduce nonlinear couplings between $A$ and $q$, which 
implies that both quantities depend in a nontrivial way
on the parameters. The values of $q$ and $A$ which minimise
$f$ are found numerically for each structure using the 
Newton-Raphson method and are then substituted into 
equations~(\ref{roll_energy}), (\ref{square_energy}) and 
(\ref{hex_energy}) to obtain the free energy density for 
different patterns.

\begin{figure}
\begin{center}
\begin{minipage}[b]{10cm}
\includegraphics[width=0.95\textwidth, angle=0]{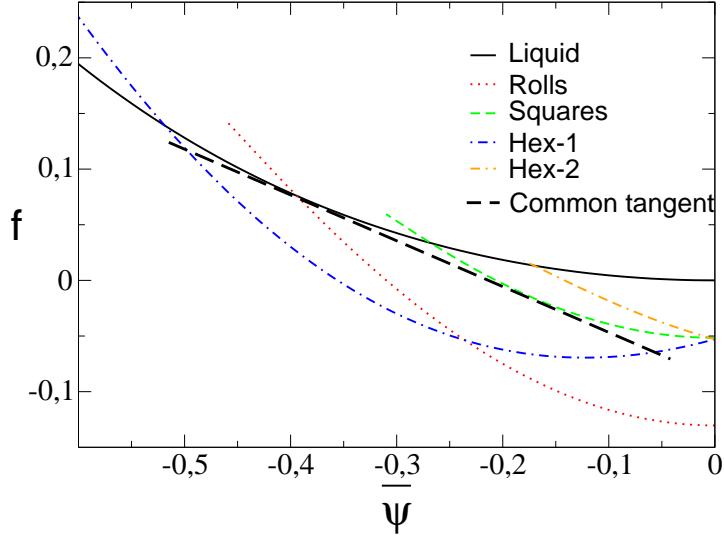}
\end{minipage}
\vskip 0.2cm
\caption{Free energy as a function of $\bar\psi$ for 
$(s_4,s_6)=(-2, 1)$, $(g_4, g_6)=(0,0)$ and $\epsilon = 0.1$. 
A common tangent line between the liquid and square free energy 
curves is drawn (thick dashed line) to illustrate the existence
of a wide coexistence region.}
\label{fig:1n2_F}
\end{center}
\end{figure}

Let us first check whether square-liquid coexistence can be
obtained using only the terms proportional to $s_4$ and $s_6$. 
In figure~\ref{fig:1n2_F} we plot the free energies of squares
and liquid for $(s_4, s_6) = (-2, 1)$, $(g_4, g_6)=(0,0)$, and 
$\epsilon = 0.1$. The common-tangent construction shows that there
indeed exists a wide coexistence region as a result of the bifurcation
being subcritical. However, the square-liquid coexistence is
metastable for these parameters, since both the roll and the hexagonal 
phases have a lower energy, as shown in figure~\ref{fig:1n2_F}.
Note that, since the
cubic term which generates hexagons breaks the symmetry between
$A$ and $-A$, there are two distinct hexagon solutions
that correspond to different values of  $A$,
which are labelled as ``Hex-1'' and ``Hex-2'' respectively in 
the graphs. To modify this ordering of the free energies, we
need to include the terms proportional to $g_4$ and $g_6$.
In the following, we give a systematic procedure to choose
appropriate values for the coefficients.

\subsection{Rolls vs. squares}
We first consider the stability of rolls and squares. As discussed
previously, choosing the coefficients of the anisotropic terms of
the right sign favours squares. However, for fixed coefficients,
the relative energies of the phases may still depend on the value 
of $\bar\psi$. Since the solid-liquid coexistence occurs near 
the density at which the solid and liquid free energy curves
intersect, stable square-liquid coexistence is only possible if the 
intersection of the square and liquid free energy curves occurs 
at a larger value of $|\bar\psi|$ than the one for rolls and liquid.
The following analysis focuses on determining the conditions
for the nonlinear coefficients for which this is the case.

The principal reciprocal lattice vectors of rolls and squares 
cannot form triads; thus, the free energy density of rolls 
and squares has the general form for a subcritical bifurcation 
given by equation~(\ref{f246}),
\begin{eqnarray}
f_S = f_L + \alpha A^2 -\beta A^4 + \gamma A^6,
\label{sub_energy}
\end{eqnarray}
where the subscript $S$ (solid) denotes rolls or squares.
The solid and liquid free energy curves intersect when the 
free energy difference of solid and liquid becomes zero, 
$\delta f \equiv f_S - f_L = \alpha A^2 -\beta A^4 + \gamma A^6=0$. 
In addition, the free energy of the solid must be minimised with 
respect to $A$, which yields 
$(\partial \delta f / \partial A) = 2 \alpha A - 4 \beta A^3 + 6\gamma A^5 = 0$. 
These two relations yield 
\begin{eqnarray}
\alpha = {\beta^2 \over 4 \gamma}.
\label{abc}
\end{eqnarray}
Comparing equations~(\ref{roll_energy}), (\ref{square_energy}) and 
(\ref{hex_energy}) to equation~(\ref{sub_energy}), we obtain
\begin{eqnarray}
\alpha = N \left(
-\epsilon + (1-q^2)^2 + 
\Gamma \bar\psi^2
\right), 
\label{apsi}
\end{eqnarray}
where
\begin{eqnarray}
\Gamma \equiv 3  +  2 g_4  q^4 -2 g_6 q^6,
\end{eqnarray}
and $N=1, 2$ and 3 for rolls, squares and hexagons, respectively.
Since the exponential growth rate of perturbations with wave number 
$q$ is proportional to $-\alpha$, in order to obtain an unconditionally 
stable liquid at large values of $|\bar\psi|$, $\alpha$ must have a 
finite lower bound which leads to
\begin{eqnarray}
\Gamma > 0.
\label{stable_liq}
\end{eqnarray}
The intersection of solid and liquid free energy curves is obtained 
by combining equations~(\ref{abc}) and (\ref{apsi}), which yields
\begin{eqnarray}
\bar\psi^{\rm int} = - {1\over \sqrt{\Gamma}}\left (
\epsilon - (1-q^2)^2 + {\beta^2 \over 4 N \gamma }
\right)^{1/2}.
\end{eqnarray}
The wave numbers $q$ that minimise the free energy of rolls 
and squares are different. However, for large values of $s_6$
this difference is small so that we can safely neglect it.
Under this assumption, the intersection point of solid and 
liquid free energy curves is solely dictated by the value of
$\beta^2/(4 N \gamma)$. The condition that
$|\bar\psi^{\rm int}_{\rm square}| > |\bar\psi^{\rm int}_{\rm roll}|$
then yields inequalities between the coefficients which are
given in the appendix. The most important result is that
this condition can only be satisfied if $g_6$ is non-zero.

\subsection{Hexagons}
For hexagons, the free energy contains a cubic term due to triadic interactions,
\begin{eqnarray}
f_{\rm hex} = f_L + \alpha A^2 + \tau A^3 - \beta A^4 + \gamma A^6,
\end{eqnarray}
where
\begin{eqnarray}
\alpha &=& 3 \left( -\epsilon + (1-q^2)^2+ \Gamma {\bar\psi}^2 \right) \nonumber \\
\tau &=& 6 \bar\psi ( \Gamma -1 ) \nonumber \\
\beta &=& -\frac{45}{2} 
          -\frac{27}{2} s_4 q^4 -84 g_4 q^4 + 264 g_6 q^6  \nonumber \\
\gamma &=& 91 s_6 q^6
\label{hex_coef}
\end{eqnarray}
In the standard PFC model, the cubic term changes the bifurcation 
from supercritical to transcritical which not only make hexagons 
the favoured phase, but also makes hexagon-liquid coexistence 
possible in the limit of a weakly first-order freezing transition.
The cubic term plays a similar role in the subcritical case: it
lowers the free energy for hexagons by an amount proportional
to $|\bar\psi|$. In order to make square symmetries favourable, 
we require the cubic coefficient $\tau$ to be small for $\bar\psi$ 
close to the square-liquid coexistence region. Then, the subcritical 
bifurcation analysis developed in the previous section still gives 
a good estimate of the intersection points for the free energy
curves of hexagons and liquid and can thus be used to evaluate
the relative stability of squares and hexagons.

\begin{figure}
\begin{center}
\begin{minipage}[b]{10cm}
\includegraphics[width=0.95\textwidth, angle=0]{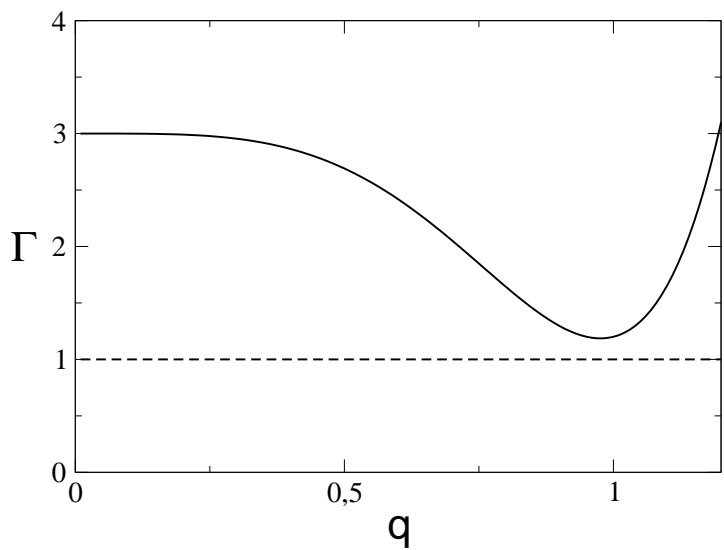}
\end{minipage}
\vskip 0.2cm
\caption{$\Gamma=3+2g_4 q^4 - 2g_6 q^6$ (solid line) for $(g_4, g_6)=(-3,-2.1)$. 
The condition for a stable liquid is satisfied since $\Gamma$ is 
positive. The cubic 
coefficient $\tau=6\bar\psi(\Gamma-1)$ is proportional to the distance 
between the solid and dashed lines. }
\label{alphas}
\end{center}
\end{figure}

This requirement on $\tau$ and the condition for a stable liquid 
are illustrated graphically in figure~\ref{alphas} for 
$(g_4, g_6)=(-3.0, -2.1)$. The condition that the liquid remains
linearly stable for large values of $\bar\psi$ requires that 
$\Gamma >0$. The cubic coefficient $\tau$ is a function of $q$ 
and proportional to the distance
between the solid and dashed lines in figure~\ref{alphas}.
We set $(s_4, s_6)=(-25, 600)$ and $\epsilon=0.001$ so that
these parameters satisfy the conditions for a subcritical bifurcation 
and for $|\bar\psi^{\rm int}_{\rm square}|> |\bar\psi^{\rm int}_{\rm roll}|$ 
as shown in 
equations~(\ref{beta_condition}) and (\ref{psi_condition}) in the appendix.
The wave number $q$ is about $0.98$ near the coexistence region, 
for which the cubic coefficient $\tau$ is small as shown in 
figure~\ref{alphas}. We can then estimate $|\bar\psi^{\rm int}_{\rm hex}|$ 
using the above analysis for a subcritical bifurcation by assuming 
$\tau \approx 0$; details can be found in the appendix. For the 
parameters chosen above, both $|\bar\psi^{\rm int}_{\rm roll}|$ and 
$|\bar\psi^{\rm int}_{\rm hex}|$ have smaller values than 
$|\bar\psi^{\rm int}_{\rm square}|$.
These parameters are used in the following numerical simulations.

\section{Comparison of analytical solutions and numerical simulations}
\label{sec:numerics}
For a conserved order parameter $\psi$, the evolution equation for 
the free energy functional shown in equation~(\ref{F_sub_cubic}) is
\begin{eqnarray}
{\partial \psi \over \partial t} &=& \nabla^2 {\delta F \over \delta \psi} \nonumber \\
&=& \nabla^2 \left(
[-\epsilon + (\nabla^2+1)^2] \psi  + \psi^3 + g_4 \psi \Delta^2 \psi^2
+ g_6 \psi \Delta^3 \psi^2 \right. \nonumber \\
& &
\left.
\mbox{} - s_4 \nabla\cdot (|\nabla \psi|^2 \nabla \psi)
-s_6 \nabla \cdot (|\nabla \psi|^4 \nabla \psi)
\right).
\label{evolution}
\end{eqnarray}
Since the main focus of this paper is the stability of crystal 
symmetries at equilibrium, we use a nonlocal globally conserved 
dynamics \cite{Mellenthin} to accelerate the search of equilibrium solutions,
\begin{eqnarray}
{\partial \psi \over \partial t} &=& = - {\delta F \over \delta \psi} + {1 \over V} \int \rmd \vec{r} {\delta F \over \delta \psi}.
\label{nAC}
\end{eqnarray}

We set the simulation parameters in two dimensions to be 
$( g_4, g_6)= (-3.0, -2.1)$, $(s_4, s_6)=(-25, 600)$, 
$\epsilon=0.001$, and the grid spacings $\Delta x$ and $\Delta y$ 
close to 0.5. Numerical simulations are carried out using 
equation~(\ref{nAC}) for rolls, squares and hexagons in one 
unit cell with initial conditions listed in equation~(\ref{ansatz}). 
The free energy is evaluated numerically using equation~(\ref{F_sub_cubic}) 
after the numerical solution reaches a steady state.
Since the wave number $q$ that minimises the free energy 
depends on $\bar\psi$, the simulation for each pattern at a fixed 
value of $\bar\psi$ is repeated with a different system size until 
the minimum of the free energy is found.

\begin{figure}
\begin{center}
\begin{minipage}[b]{10cm}
\includegraphics[width=0.95\textwidth, angle=0]{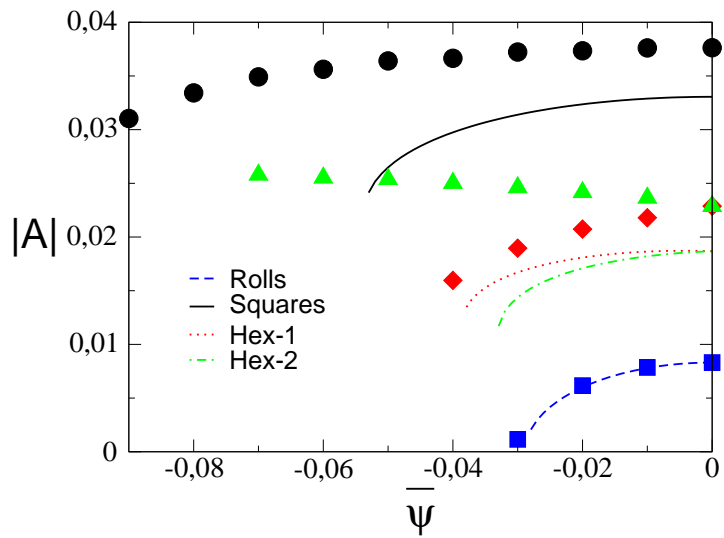}
\end{minipage}
\vskip 0.2cm\caption{Comparison of the magnitude of $A$ that minimises 
the free energy as a function of $\bar\psi$ for different patterns. 
The analytical solutions are plotted as lines. The numerical simulations 
for rolls, squares, Hex-1 and Hex-2 are plotted as squares, circles, 
diamonds and triangles, respectively.}
\label{amps}
\end{center}
\end{figure}

A comparison of the amplitudes of the principal reciprocal lattice 
vectors obtained from the numerical simulations and the corresponding 
analytical solution is shown in figure~\ref{amps}.
In the numerical simulations, the amplitudes of the principal 
reciprocal lattice vectors are computed using the Fourier transform.
For rolls, the one-mode approximation and the numerical simulations 
are in good agreement. This is reasonable because the next-nearest 
reciprocal lattice vectors of rolls are $\langle 20\rangle$, which 
is far from the $\langle 10\rangle$ principal reciprocal lattice vectors. 
This makes 
them difficult to excite, and the one-mode approximation is quite accurate. 
In contrast, for squares and hexagons, the higher-order reciprocal 
lattice vectors are closer to the principal reciprocal lattice 
vectors (e.g., $\langle 11\rangle$ for squares), and thus the 
one-mode approximation and the numerical simulations do not agree 
as well. Thus, an analytical calculation including higher-order 
modes is required to give more accurate predictions.
Nevertheless, the one-mode approximation gives a good qualitative 
prediction of the relative values of the free energies at the end point 
of the solid solution branches, as well as for the intersection points  
of solid and liquid free energy curves.

\begin{figure}
\begin{center}
\begin{minipage}[b]{11cm}
\includegraphics[width=0.95\textwidth, angle=0]{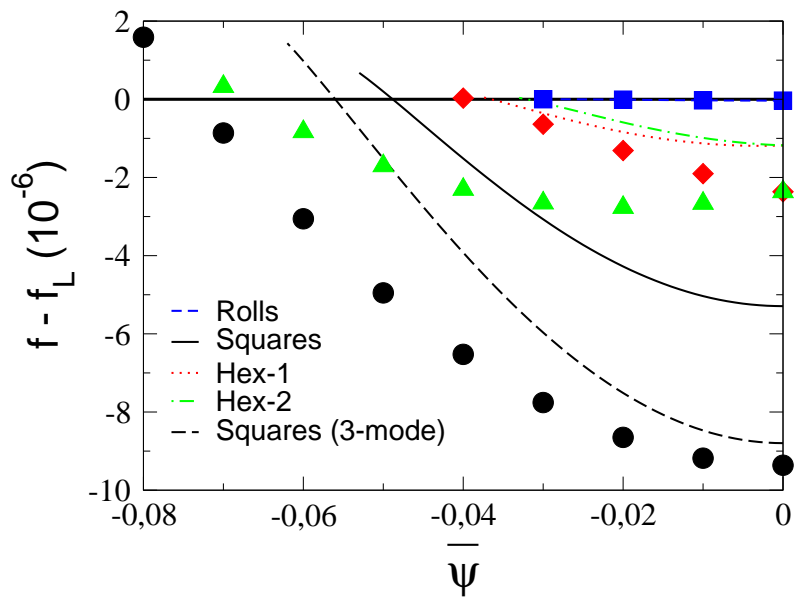}
\end{minipage}
\vskip 0.2cm\caption{Comparison of the free energy densities as 
a function of $\bar\psi$ for different patterns. Symbols and 
lines as in figure~\ref{amps}.}
\label{energies}
\end{center}
\end{figure}

The free energy difference $f-f_L$ is plotted in figure~\ref{energies}.
This difference is so small that it would be difficult to distinguish 
the free energy curves if they were plotted as in figure~\ref{fig:1n2_F}.
The one-mode approximation and numerical simulations show the same 
ordering of the intersection points of solid and liquid free energy 
curves, as in  figure~\ref{amps}, namely 
$|\bar\psi^{\rm int}_{\rm square}| > |\bar\psi^{\rm int}_{\rm hex}| > |\bar\psi^{\rm int}_{\rm roll}| $.
In addition, for the parameters chosen above, rolls and hexagons are 
metastable states and the square lattice is the ground state for a 
wide range of $|\bar\psi|$ until it loses its stability to liquid 
at $|\bar\psi|=0.073$. To illustrate the influence of triadic interactions
on the free energy calculation, a three-mode approximation that considers 
$\langle 10\rangle$, $\langle 11\rangle$ and $\langle 12\rangle$ reciprocal 
lattice vectors for the square lattice is computed and shown in 
figure~\ref{energies}. 

\begin{figure}
\begin{center}
\begin{minipage}[b]{12cm}
\includegraphics[width=0.95\textwidth, angle=0]{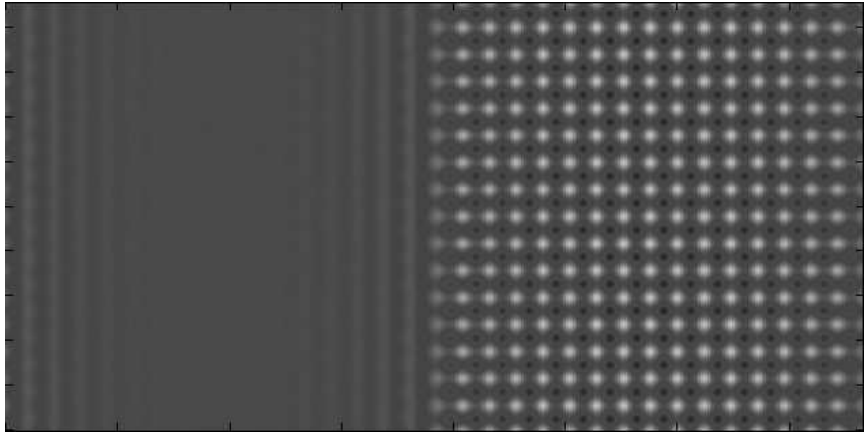}
\end{minipage}
\vskip 0.2cm\caption{Square-liquid coexistence simulated using the
free energy functional shown in equation~(\ref{F_sub_cubic}).}
\label{energies2}
\end{center}
\end{figure}

To simulate solid-liquid coexistence, we set the simulation parameters to 
$\Delta x= \Delta y = 0.53$ on a system of size $L_x=384 \Delta x$ and 
$L_y = 192 \Delta y$ with periodic boundary conditions.
The initial condition for the density is
\begin{eqnarray}
\psi = {1 \over 2} \left( 1+ \tanh{(x- {L_x \over 2}) } \right)
\psi_S(\vec{r}) 
+ {1\over 2} \left( 1 - \tanh{(x-{L_x \over 2} ) } \right) \bar\psi_L,
\end{eqnarray} 
where $\psi_S$ is the one-mode approximation of square lattices 
as shown in equation~(\ref{ansatz}),
and $\bar\psi_L$ is the constant density of the liquid.
The average densities of solid and liquid are chosen to be close to 
$\bar\psi^{\rm int}_{\rm square}$, which is determined numerically.
The initial amplitude of the square lattice is set to the value obtained 
from the steady state square lattice simulations at the same average density.
Uniformly distributed random fluctuations of the magnitude of half the 
square pattern amplitude are applied initially to examine the 
stability of the square pattern. The simulations show that the 
square-liquid coexistence is stable against the initial random 
fluctuations, and the equilibrium solid-liquid coexistence is 
shown in figure~\ref{energies2}. Furthermore, it can be seen
in figure ~\ref{energies2} that the square-liquid interface 
displays different spatial decay rates for the different density waves
as predicted by the classical density functional theory of freezing.
In particular, the density wave of $\langle 10 \rangle$ decays more
slowly into the liquid than the density wave of $\langle 01 \rangle$ 
for the $\{10\}$ interface. The directional dependence of the 
spatial decay rate of density waves into the liquid was shown to be 
a main determinant of the anisotropy of interface properties, as 
discussed in \cite{Wu07, Wu06}.

\section{Elastic constants}
\label{sec:elastic}
We briefly investigate here the elastic properties of our model.
This is important because the analysis of the two-mode model
presented in \cite{Wu10} predicts that the shear modulus for
square patterns is zero in the limit of small $\epsilon$ 
if only the mode corresponding to the principal 
reciprocal lattice vectors $\langle 10\rangle$
is active. Only the addition of the second mode corresponding
to the second reciprocal lattice vectors $\langle 11\rangle$
ensures a finite and positive shear modulus. Since our model
is based on a single unstable mode, it is important to evaluate
the shear modulus.

To determine the elastic constants, we follow the lines of
earlier work \cite{Elder04,Wu10} and consider perturbations
of the density field given by equation (\ref{ansatz}). For 
shear, bulk and deviatoric deformations, we use
\begin{eqnarray}
\psi_{\rm shear} = \bar\psi + 2 A \left( \cos{q (x+\xi y)} + \cos{qy} \right),
\end{eqnarray}
\begin{eqnarray}
\psi_{\rm bulk} = \bar\psi + 2A \left( 
\cos{qx\over (1+\xi)} 
+\cos{qy\over (1+\xi)} 
\right),
\end{eqnarray}
and
\begin{eqnarray}
\psi_{\rm deviatoric} = \bar\psi + 2A \left( 
\cos{qx\over (1+\xi)} 
+\cos{qy\over (1-\xi)} 
\right),
\end{eqnarray}
respectively, where $\xi$ is the strain. These expressions
are then inserted in the free energy functional (\ref{F_sub_cubic}), 
which can be explicitly evaluated in terms of $q$, $A$, and $\xi$.
We set the wave number $q$ equal to the wave number of the 
reference (unstrained) state, and determine $A$ by minimising
the free energy of the strained state for each value of $\xi$
numerically using again the Newton-Raphson method.
This value is then used to evaluate the free energy density,
as a function of $\xi$. The difference with the free energy
of the unstrained state, $\Delta f$, can be related to the
elastic constants. To lowest order in $\xi$, we have
\begin{eqnarray}
\Delta f_{\rm shear} &=& {C_{44} \over 2} \xi^2 \nonumber \\
\Delta f_{\rm bulk} &=& ({C_{11} + C_{12} }) \xi^2 \nonumber \\
\Delta f_{\rm deviatoric} &=& ({C_{11} - C_{12} }) \xi^2. 
\end{eqnarray}

We have evaluated the free energies of the strained states
for the same parameters as the numerical simulations shown
previously, except that we set $\bar\psi=0$ for simplicity.
The results indeed display the expected quadratic behaviour
for $\xi\ll 1$, and the elastic constants are, in our
dimensionless units,
\begin{eqnarray}
C_{11} &=& 7.8 \times 10^{-3}, \nonumber \\
C_{12} &=& 1.1 \times 10^{-3}, \nonumber \\
C_{44} &=& 1.63 \times 10^{-4}.
\end{eqnarray}
The shear modulus is $\mu = C_{44} = 1.63 \times 10^{-4}$, and the 
(two-dimensional) bulk modulus 
$B = ({C_{11} + C_{12} })/2 = 3.85 \times 10^{-3}$.
The ratio of the bulk modulus and the shear modulus is about 24.

Thus, for the parameters chosen here the square patterns generated by our 
model are rather ``soft'' with respect to shear deformation, but have a 
perfectly well-defined finite shear modulus. Several remarks can help to
understand this fact. A shear deformation corresponds, in reciprocal
space, to a change of the angle between reciprocal lattice
vectors, whereas their length remains unchanged to first
order in $\xi$. Since we
have found that the free energy depends on this angle if
anisotropic terms are present according to equation 
(\ref{anisotropy}), this directly gives a finite
contribution to the shear modulus. Note, however, that
this is not the only contribution. Indeed, since the
change in the angle modifies the strength of the nonlinearities
which saturate the density waves, the amplitude $A$
also depends on the deformation, which modifies the
contributions of all the other terms in the functional.
Furthermore, the nature of the bifurcation also comes into play. 
For a subcritical bifurcation, $A$ can be appreciable even 
for small $\epsilon$, which means that the anisotropic terms, 
though proportional to $A^4$ or $A^6$, may be comparable to 
the quadratic contribution which determines the elastic
constants in the standard PFC model.

Note that the above calculation of the elastic constants
uses the one-mode approximation, which is not very 
accurate in view of the results shown in figure~\ref{amps}. 
However, it is sufficient for our main purpose here, which was 
to demonstrate that the shear modulus is finite. For a more quantitative
evaluation of the elastic constants, multi-mode calculations
or direct numerical computations are mandatory. Furthermore,
we have evaluated the elastic constants here only for one specific
set of coefficients. How they depend on the various parameters
in the free energy functional is an interesting subject for
further investigations.

\section{Conclusions}
\label{sec:conclusions}
We have presented a general method of constructing a free energy 
functional for the PFC model that exhibits solid-liquid coexistence 
with desired crystal symmetries. We have demonstrated that the 
crystal symmetries in the PFC model can be controlled through 
additional nonlinear terms. In particular, we have examined 
the influence of nonlinearities such as $g_{2n} \psi^2 \Delta^n \psi^2$ 
and $s_{2m}|\nabla\psi|^{2m}$ on crystal symmetries. Besides the
squares that have been investigated here, other structures can
potentially be obtained by using terms with $n\geq 4$, since 
these terms contain higher order harmonics that can be used
to favour well-defined angles between density waves.

We have also presented a systematic procedure for constructing 
free energy functionals that exhibit square-liquid coexistence 
as the ground state. This requires (i) a subcritical bifurcation
from the liquid to the square state, and (ii) a suitable 
combination of the nonlinear coefficients which favours squares
over rolls and, at the same time, makes triadic resonances
small enough to avoid hexagons. Our analytical results obtained
in the one-mode approximation are borne out qualitatively by 
numerical simulations: in both simulations and analytics the 
square lattice is the ground state until it loses its 
stability to the liquid at large $|\bar\psi|$. The one-mode 
approximation, however, does not predict accurately the 
amplitude and free energy for square lattices. A multi-mode 
approximation is needed to quantitatively describe the square 
symmetry since the principal lattice vectors of the square 
lattice are strongly coupled to higher order modes. 
Nevertheless, the one-mode analysis provides a good guideline 
for determining coefficients of nonlinearities to obtain 
the desired crystal symmetries.

We have obtained stable square-liquid coexistence, and the solid
exhibits a finite shear modulus, which makes this model suitable
for simulations. A drawback is that several nonlinear terms are 
needed, and that a delicate balance between the coefficients needs to be
respected. We have only tested one specific set of coefficients,
but we expect that the properties of the model (elastic constants,
interfacial properties) can be varied by exploring the parameter
space of the various coefficients. It should also be noted that
our survey of potential nonlinear terms is by no means complete.
However, the relation between the coefficients and the resulting
model properties
is highly non-trivial, which implies that the process of finding 
a set of coefficients that yield a match with some desired materials 
properties is likely to be cumbersome. Furthermore, our final equation
of motion (\ref{evolution}) is of 8th order in space, which is only 
two orders lower than the one of the two-mode model \cite{Wu10}.

A highly interesting perspective is the extension of this work
to three dimensions. The calculation of the anisotropy performed
in section \ref{sec:model} remains valid in three dimensions,
which means that it is straightforward to obtain simple cubic
structures. Furthermore, since it has been found that the
free energies of bcc, fcc, and hcp phases in three dimensions
in the standard PFC model are not very different for a certain 
range of $\epsilon$ \cite{Tegze09,Toth10,Jaatinen10}, the addition
of anisotropic terms may be used to control their relative stability.

\ack
This work was supported by joint funding under EU STRP 016447 
MagDot and NSF DMR Award No. 0502737.

\appendix

\section{Inequalities for the coefficients in the free energy functional}

Here, we give the details concerning the calculation of the
intersection points between liquid and solid free energies
and the resulting conditions for the coefficients in the
free energy functional. We start with rolls and squares. 
As outlined in section~3.2, we need to compare the values
of $\beta^2/(4N\gamma)$, where $\beta$, and $\gamma$ are 
the coefficients of equation (\ref{sub_energy}) for a 
generic subcritical bifurcation.
Comparing equations~(\ref{roll_energy}) and (\ref{square_energy}) 
to equation~(\ref{sub_energy}), we obtain 
the coefficients $\beta$ and $\gamma$ for rolls and squares, 
\begin{eqnarray}
\beta_{\rm roll} &=&  
{1\over 2} (-3-3 s_4 q^4 + \lambda_1 \Xi ), \nonumber \\
\gamma_{\rm roll} &=& {10 \over 3} s_6 q^6, \\
\beta_{\rm square} &=&  
  (-9-5 s_4 q^4 + \Xi), \nonumber \\
\gamma_{\rm square} &=& {56 \over 3} s_6 q^6,
\end{eqnarray}
where 
\begin{eqnarray}
\Xi &\equiv& - 32g_4 q^4 +96 g_6 q^6  , \nonumber \\
\lambda_1 &\equiv& {(  - 16g_4 q^4 + 64 g_6 q^6  )/   \Xi }.
\end{eqnarray}
The subcritical bifurcation requires the quartic coefficient to be negative,
which according to equation~(\ref{f246}) implies
\begin{eqnarray}
\beta_{\rm square} > 0.
\end{eqnarray}
This determines an upper bound for $s_4$,
\begin{eqnarray}
s_4 q^4 < {{-9+\Xi} \over 5 }.
\label{beta_condition}
\end{eqnarray}
The condition that the intersection of the square and liquid 
free energy curves occurs at a higher value of $|\bar\psi|$ than
that of rolls and liquid (i.e., 
$|\bar\psi^{\rm int}_{\rm square}| > |\bar\psi^{\rm int}_{\rm roll}|$), yields
\begin{eqnarray} 
{\beta^2_{\rm square} \over 8 \gamma_{\rm square} } > {\beta^2_{\rm roll} \over 4 \gamma_{\rm roll} }, 
\end{eqnarray} 
which can be reduced to the inequality
\begin{eqnarray}
99 - 25 \Xi + 42 \lambda_1 \Xi - \sqrt{\Omega_1}<s_4 q^4 < 99 - 25 \Xi + 42 \lambda_1 \Xi + \sqrt{\Omega_1},
\label{psi_condition}
\end{eqnarray}
where
\begin{eqnarray}
\Omega_1 &=& 
70 (5 \lambda_1 \Xi - 3 \Xi + 12)^2.
\end{eqnarray}
The solution of $s_4 q^4$ exists only if the upper bound of 
$s_4 q^4$ obtained from equation~(\ref{beta_condition}) is 
greater than the lower bound obtained from equation~(\ref{psi_condition}), 
which yields
\begin{eqnarray}
(-9+\Xi)/5 > 99 - 25 \Xi + 42 \lambda_1 \Xi - \sqrt{70}\; |5 \lambda_1 \Xi - 3 \Xi + 12|.
\end{eqnarray}
The above inequality holds if (a) $\lambda_1 < 3/5$ and 
$\Xi > 12/(3-5 \lambda_1)$ or (b) $\lambda_1 > 3/5$ and  $\Xi < 12/(3-5 \lambda_1)$.
It is helpful to rewrite the expression of $\lambda_1$ as 
\begin{eqnarray}
\lambda_1 = {3 \over 5} + {4\over 5} \cdot 
{{ 4 g_4 q^4 +8 g_6 q^6 }
\over 
{ \Xi }
}.
\label{lambda}
\end{eqnarray}
For the case that $g_6 = 0$, we have $\lambda_1 =3/5-1/10<3/5$ and 
the solution of $s_4$ only exists if $\Xi > 12/(3-5 \lambda_1) > 0$.
However, this requires $g_4$ to be negative which contradicts 
the condition for stable liquid shown in equation~(\ref{stable_liq}).
Thus it is essential to include the nonlinear term $\psi^2 \Delta^3 \psi^2$ 
in the free energy functional so that the free energy functional 
exhibits the desired properties of (i) stable liquid, (ii) subcritical 
bifurcation, and (iii) $|\bar\psi^{\rm int}_{\rm square}| > |\bar\psi^{\rm int}_{\rm roll}|$.

For hexagons, the coefficients $\alpha$, $\beta$, and $\gamma$
are given by equation~(\ref{hex_coef}). The coefficient $\tau$ of 
the cubic term is assumed to be small, such that this term can be 
neglected. Then, the condition that 
$|\bar\psi^{\rm int}_{\rm square}|>|\bar\psi^{\rm int}_{\rm hex}|$ requires
\begin{eqnarray}
{\beta^2_{\rm square} \over 8 \gamma_{\rm square} } > {\beta^2_{\rm hex} \over 12 \gamma_{\rm hex} }, 
\end{eqnarray}
yielding the inequality
\begin{eqnarray}
-45 + 65 \Xi + 24 \lambda_2 \Xi - \sqrt{\Omega_2} < s_4 q^4 <
-45 + 65 \Xi + 24 \lambda_2 \Xi + \sqrt{\Omega_2},
\end{eqnarray}
where 
\begin{eqnarray}
\lambda_2 \equiv (84 g_4 q^4 - 264 g_6 q^6) / \Xi,
\end{eqnarray}
and
\begin{eqnarray}
\Omega_2 = {52 \over 9} \left( 
10 \lambda_2 \Xi + 27 \Xi -18
\right)^2.
\end{eqnarray}
Together with equation~(\ref{beta_condition}), we find that 
either (a) $\lambda_2 > -27/10$ and $\Xi > 18/(27+15 \lambda_2)$ or
(b) $\lambda_2 < -27/10$ and $\Xi < 18/(27+15 \lambda_2)$
has to be fulfilled in order to make 
$|\bar\psi^{\rm int}_{\rm square}| > |\bar\psi^{\rm int}_{\rm hex}|$.


\section*{References}
\begin{thebibliography}{10}
\bibitem{Elder02} Elder K R, Katakowski M, Haataja M and Grant M 2002 {\it Phys. Rev. Lett.} {\bf 88} 245701
\bibitem{Elder04} Elder K R and Grant M 2004 {\it Phys. Rev. E} {\bf 70} 051605
\bibitem{SH}  Swift J and  Hohenberg P C 1977 {\it Phys. Rev. A} {\bf 15} 319 
\bibitem{Cross93} Cross M C and Hohenberg P C 1993 {\it Rev. Mod. Phys.} {\bf 65} 851
\bibitem{Elder07} Elder K R,  Provatas N,  Berry J,  Stefanovic P and Grant M 2007 {\it Phys. Rev. B} {\bf 75} 064107 
\bibitem{Wu07} Wu K-A and Karma A 2007 {\it Phys. Rev. B} {\bf 76} 184107
\bibitem{SvT09} van Teeffelen S, L\"owen H, Backofen R and Voigt A 2009 {\it Phys. Rev. E} {\bf 79} 051404
\bibitem{Ramakrishnan79} Ramakrishnan T V and Youssouff M 1979 {\it Phys. Rev. B} {\bf 19} 2775
\bibitem{Singh91} Singh Y 1991 {\it Phys. Reports} {\bf 207} 351
\bibitem{Tegze09} Tegze G, Gr\'an\'asy L T\'oth G I, Podmaniczky F, Jaatinen A, Ala-Nissila T and Pusztai T 2009 {\it Phys. Rev. Lett.} {\bf 103} 035702
\bibitem{Toth10} T\'oth G I, Tegze G, Pusztai T, and Gr\'an\'asy L 2010 this volume
\bibitem{Jaatinen10} Jaatinen A and Ala-Nissila T 2010 this volume
\bibitem{Lifshitz97} Lifshitz R and Petrich D M 1997 {\it Phys. Rev. Lett.} {\bf 79} 1261
\bibitem{Wu10} Wu K-A, Adland A and Karma A 2010 Phase-field crystal model for fcc ordering {\it Preprint} arXiv:1001.1349
\bibitem{Greenwood10} Greenwood M, Provatas N and Rottler J 2010 Free energy functionals for efficient phase field crystal modeling of structural phase transformations {\it Preprint} arXiv:1002.3185 
\bibitem{Busse78} Busse F H 1978 {\it Rep. Prog. Phys.} {\bf 41} 1930
\bibitem{Gertsberg81} Gertsberg V L and Sivashinski G I 1981 {\it Prog. Theor. Phys.} {\bf 66} 1219
\bibitem{Bestehorn84} Bestehorn M and Haken H 1984 {\it Z. Phys. B} {\bf 57} 329
\bibitem{Herrero94} Herrero H, P\'erez-Garcia C and Bestehorn M 1994 {\it Chaos} {\bf 4} 15
\bibitem{Kubstrup96} Kubstrup C, Herrero H and P\'erez-Garcia C 1996 {\it Phys. Rev. E} {\bf 54} 1560
\bibitem{Matsushita98} Matsushita N and Ohta T 1998 {\it J. Phys. Soc. Japan} {\bf 67} 1973
\bibitem{Enomoto01} Enomoto Y, Oba K, Hayase Y and Ohta T 2001 {\it J. Phys. Soc. Japan} {\bf 70} 2939
\bibitem{Boyer05} Boyer D and Romeu D 2005 {\it Int. J. Mod. Phys. B} {\bf 19} 4047
\bibitem{Sakaguchi97} Sakaguchi H and Brand H R 1997 {\it Europhys. Lett.} {\bf 38} 341
\bibitem{Crawford99} Crawford C and Riecke H 1999 {\it Physica D} {\bf 129} 83
\bibitem{Mellenthin} Mellenthin J, Karma A and Plapp M 2008 {\it Phys. Rev. B} {\bf 78} 184110
\bibitem{Wu06} Wu K-A, Karma A, Hoyt J J and Asta M 2006 {\it Phys. Rev. B} {\bf 73} 094101
\end{thebibliography}
\end{document}